\documentclass[10pt]{article}
\usepackage{titlesec}
\usepackage{amsmath,esdiff}
\usepackage{amssymb}
\usepackage{authblk}

\usepackage[normalem]{ulem}
\usepackage{bbold}
\usepackage[colorlinks=true,linkcolor=blue,citecolor=teal,urlcolor=teal]{hyperref}%
\usepackage{geometry}
\usepackage{setspace}
\usepackage{cite}
\usepackage{flushend}
\usepackage{mathrsfs}
\usepackage{hyperref}
\usepackage{graphicx}
\usepackage{cancel}
\usepackage{array,esint}
\usepackage[most]{tcolorbox}
\graphicspath{}
\titleformat{\section}{\fontsize{12}{12}\bfseries}{\thesection}{1em}{}
\twocolumn
\begin{document}
\twocolumn[\begin{@twocolumnfalse}
\title{\textbf{Phase transition structure and breaking of universal nature of central charge criticality in a Born-Infeld \textit{AdS} black hole}}
\author{\textbf{Neeraj Kumar${}^{a*}$, Soham Sen${}^{a\dagger}$ and Sunandan Gangopadhyay${}^{a\ddagger}$}}
\affil{{${}^a$ Department of Theoretical Sciences}\\
{S.N. Bose National Centre for Basic Sciences}\\
{JD Block, Sector III, Salt Lake, Kolkata 700 106, India}}
\date{}
\maketitle
\begin{abstract}
In this paper we have considered the thermodynamics of  a Born-Infeld $AdS$ black hole using inputs from the dual boundary  field theory. Here, we have varied the cosmological constant $\Lambda$ and the Newton's gravitational constant $G$ along with the Born-Infeld parameter $b$ in the bulk. A novel universal critical behaviour of the central charge (occurring in the boundary conformal field theory) in extended black hole thermodynamics for charged black holes has been recently observed \cite{mann1}, and we have extended this study to Born-Infeld $AdS$ black holes. The Born-Infeld parameter has the dimension of inverse length, therefore, when considered in the first law of thermodynamics of the bulk in the mixed form which includes the central charge of the boundary conformal field theory, it modifies the thermodynamic volume and the chemical potential (which are conjugate to pressure and central charge respectively).  We observe that due to this inclusion of the Born-Infeld non-linearity in this analysis, the universal nature of the critical value of the central charge observed in \cite{mann1} breaks down. We also observe an interesting behaviour of the free energy of the black hole with Hawking temperature due to the variations in both the central charge and the Born-Infeld parameter. It is also observed in our analysis that for a sufficiently small value of the Born-Infeld parameter (small value of this parameter has more prominent non-linear effects), there exists a critical value of the temperature below which no black hole can exist.
\end{abstract}
\end{@twocolumnfalse}]
\section{Introduction}
\noindent\let\thefootnote\relax\footnote{{}\\
{$*$neerajkumar@bose.res.in}\\
{$\dagger$sensohomhary@gmail.com, soham.sen@bose.res.in}\\
{$\ddagger$sunandan.gangopadhyay@gmail.com}}
General theory of relativity and quantum mechanics are the two most revolutionary theories of the previous century, describing the two extreme length scales of our universe. Further, the unification of gravity and quantum mechanics has turned out to be a very complicated problem. In the seminal works\cite{Hawking,Hawking2,Hawking3}, Stephen Hawking showed that when we consider quantum effects in curved spacetime, a black hole emits radiation. The formulation of black hole thermodynamics\cite{Hawking,Hawking2,Hawking3,Bekenstein,Bekenstein2}, Hawking radiation\cite{Hawking,Hawking2,Hawking3}, particle emission from black holes\cite{Page,Page2,Page3}, and acceleration radiation\cite{Unruh,Fulling21, Davies, DeWitt, Unruh2, Muller, Vanzella, Higuichi, Ordonez1, Ordonez2, Ordonez3, Ordonez4}, generated a link between gravitation, geometry and thermodynamics. The thermodynamics of black holes has become a field of extreme importance in recent times. The understanding of asymptotically anti-de Sitter ($AdS$) black holes, via the $AdS/CFT$ correspondence\cite{mald}, provides a deep insight into the dual conformal field theory ($CFT$) at finite temperature. When in thermal equilibrium with the Hawking radiation, $AdS$ black holes show interesting phase transition behaviour, namely, first order Hawking-Page phase transition\cite{Hawking4,Hawking5}, first order phase transition in charged Reissner-Nordstr$\ddot{\text{o}}$m $AdS$ black holes\cite{Chamb,Chamb2},  and small to large black hole phase transitions\cite{Cvetic,Mann2,RMSBSG,NKSG1,NKSBSG2}.

\noindent In recent times, there has been an upsurge in considering the variation of the cosmological constant ($\Lambda$) in the first law of black hole thermodynamics\cite{mann1,Mann2, MM, kastor, Dolan, Dolan2, Dolan3, Cvetic2, LuPang}. In standard black hole thermodynamics, black hole parameters are varied in a fixed $AdS$ background with $\Lambda$ being kept fixed. Now considering variations in the fundamental constants, like Newton's gravitational constant, cosmological constant, gauge coupling constants lead us to more fundamental theories\cite{Gibbons, Mann3}. It has been argued in \cite{kastor} that the first law of black hole thermodynamics becomes inconsistent when viewed at the level of the Smarr relation\cite{Smarr} in the presence of a fixed cosmological constant. Therefore, to counter this issue, $\Lambda$ is included in the first law \cite{kastor} as a thermodynamic variable, with a negative cosmological constant being identified with a positive thermodynamic pressure. In case of a Born-Infeld \cite{Born_Inf} black hole, the Born-Infeld parameter $b$ must be varied in the first law of the black hole thermodynamics in order to make it consistent with the Smarr relation\cite{Rash,Breton,Huan}. This consideration has opened up a new direction of investigation which is the extended thermodynamic behaviour of $AdS$ black holes with novel phase transition structures in contrast to standard black hole thermodynamics\cite{ext1,ext2,gunasekaran,ext3,ext4,ext5}. A detailed analysis on extended phase space thermodynamics for charged and rotating black holes was carried out in \cite{gunasekaran}. Interesting insight about this extended black hole chemistry\cite{revmann1,bhc} can be further made using the gauge/gravity duality. It has been a bit tricky to find a holographic insight into black hole chemistry\cite{kostar2,Dolan1,Kastor,Zhang,Zhang2,Dolan22,McCarthy}. A simple argument indicates that the relation of the first law (of the Bulk) to the thermodynamics of the holographic field theory is not easy\cite{karch,Sinamuli,Visser}. The reason is that varying the cosmological constant would in turn mean varying the central charge and the CFT volume of the boundary field theory\cite{kostar2}. This would in turn imply that the CFT gets changed, and therefore the identification of the volume conjugate to the thermodynamic pressure in the bulk would become problematic. The way out of this problem is to vary $G$ along with the cosmological constant $\Lambda$ so that the central charge of the dual CFT remains fixed. This leads to a mixed form of the extended first law of thermodynamics which enables one to identify the appropriate thermodynamic volume and chemical potential. Following this approach we write down the mixed form of the first law of thermodynamics in the bulk for Born-Infeld $AdS$ black hole.

\noindent A duality between holographic and bulk thermodynamics begins from the relation between the central charge of the CFT and the parameters on the gravity side, namely, the $AdS$ radius and Newton's gravitational constant. A free energy analysis of the bulk thermodynamics carried out in \cite{mann1} revealed that there is a critical central charge above which the free energy has a swallowtail behaviour for any $P$ which implies a universal behaviour. Our aim is to look at this universality in the case of Born-Infeld $AdS$ black hole. This would give us an idea about non-linear effects in these type of phase transitions.

\noindent The organization of this paper is given as follows. In the next section, we have provided a brief introduction of the Born-Infeld $AdS$ black holes in 4 spacetime dimensions and calculated the Hawking temperature for the same. In section 3, we have calculated the mixed form of the extended first law of  black hole thermodynamics in the bulk and obtained the modified thermodynamic variables using the modified Smarr relation. Next we have calculated the critical value of the central charge in order to inspect its behaviour and then we have plotted the different phase transition structures based on the variation of the central charge and the Born-Infeld parameter respectively.  
\section{Born-Infeld \textit{AdS} black hole and Hawking temperature}
\noindent We start with a brief review of the Born-Infeld $AdS$ black hole. The solution with a cosmological constant was first constructed in \cite{tanay}. The action in ($3+1$) dimensional spacetime reads 
\begin{equation}\label{1.1}
S=\dfrac{1}{16\pi G}\int d^{4}x\sqrt{-g}\left[R-2\Lambda+L(F)\right]~
\end{equation}
where the Born-Infeld part of the Lagrangian is of the form
\begin{equation}\label{1.2}
L(F)=4b^2\left(1-\sqrt{1+\dfrac{F^{\mu\nu}F_{\mu\nu}}{2b^2}}\right) ~.
\end{equation}
 $\Lambda$ is the cosmological constant and $b$ is the Born-Infeld parameter.
The above action admits the following black hole solution \cite{tanay}
\begin{equation}\label{1.3}
ds^2=-f(r)dt^2+\dfrac{dr^2}{f(r)}+r^2d\Omega^2
\end{equation}
where 
\begin{equation}\label{1.4}
\begin{split}
f(r)&=1-\dfrac{2GM}{r}+\dfrac{r^2}{l^2}+\dfrac{2b^2r^2}{3}\left(1-\sqrt{1+\dfrac{GQ^2}{b^2r^4}}\right) \\
&+\frac{4GQ^2}{3r^2}\, \,_2F_1\left[\frac{1}{4},\frac{1}{2},\frac{5}{4},-\dfrac{GQ^2}{b^2r^4}\right]~.
\end{split}
\end{equation}
In eq.(\ref{1.4}), $M$ is the mass of the black hole, $Q$ is the charge, $l$ is the AdS radius and $\,_2F_1$ is the Gauss hypergeometric function. In the limit $b\rightarrow\infty$, the above metric reduces to the $AdS$ Reissner-Nordst$\ddot{\text{o}}$rm black hole. The cosmological constant $\Lambda$ in terms of the AdS radius $l$ is $\Lambda=-\frac{3}{l^2}$ in $(3+1)$-dimensions.\

\noindent The event horizon of the black hole is obtained from the relation $f(r_+)=0$. This relates the mass $M$ in terms of the horizon radius ($r_+$) as
\begin{equation}\label{1.6}
\begin{split}
M&=\dfrac{r_+}{2G}+\dfrac{r_+^3}{2G l^2}+\dfrac{b^2r^3_+}{3G}\left(1-\sqrt{1+\dfrac{GQ^2}{b^2r_+^4}}\right)\\
&+\dfrac{2Q^2}{3r_+}\,_2F_1\left[\frac{1}{4},\frac{1}{2},\frac{5}{4},-\dfrac{GQ^2}{b^2r_+^4}\right]~.
\end{split}
\end{equation}
We can obtain the Hawking temperature of the black hole from eq.(\ref{1.4}) as follows
\begin{equation}\label{1.7}
\begin{split}
T&=\dfrac{1}{4\pi}\dfrac{\partial f}{\partial r}\bigg\vert_{r=r_+}\\&=\dfrac{1}{4\pi}\left[\dfrac{1}{r_+}+\dfrac{3r_+}{l^2}+2b^2r_+\left(1-\sqrt{1+\dfrac{GQ^2}{b^2r_+^4}}\right)\right]~.
\end{split}
\end{equation}
We will use eq.(s)(\ref{1.6},\ref{1.7}) to calculate the critical value of the central charge of the Born-Infeld $AdS$ black hole in section 4. In the next section, we will try to find the modified form of the first law of the black hole thermodynamics in $D$ dimensions.
\section{The modified first law of thermodynamics}
\noindent It has been realised recently that a negative cosmological constant can induce a positive thermodynamic pressure\cite{mann1,ext2}. This led to the inclusion of the cosmological constant in the extended thermodynamic phase space. Here, we take another step forward by treating the Born-Infeld parameter $b$ as a thermodynamical variable as we would like to see the effects of this parameter on the first law of thermodynamics. The pressure of the black hole in terms of the cosmological constant and the Newton's gravitational constant can be written in the following  form\cite{kastor}

\begin{equation}\label{1.8}
P=-\dfrac{\Lambda}{8\pi G}~.
\end{equation}
From the above equation, we observe that with the variation in the cosmological constant, the thermodynamic pressure of the black hole also changes. Now, the cosmological constant in general $D$-dimensions in terms of $AdS$ radius can be written as
\begin{equation}\label{1.9}
\Lambda=-\frac{(D-1)(D-2)}{2l^2}~.
\end{equation}
It can be seen from eq.(s)(\ref{1.8},~\ref{1.9}) that a negative cosmological constant induces a positive thermodynamic pressure.

\noindent In natural units, the Bekenstein-Hawking entropy takes the form \cite{Hawking,Hawking2,Hawking3,Bekenstein,Bekenstein2}
\begin{equation}\label{1.10}
S=\frac{A}{4G}
\end{equation}
where $A$ is the area of the black hole. In terms of the surface gravity $\kappa$, the Hawking temperature of the black hole takes the form 
\begin{equation}\label{1.11}
T=\frac{\kappa}{2\pi}~.
\end{equation} 
The mass of the black hole in the extended thermodynamic phase space\cite{kastor} can be thought of as enthalpy rather than internal energy. Therefore, the most general form of the first law of thermodynamics for a black hole with surface gravity $\kappa$, charge $Q$, angular momentum $J$ and area $A$ can be written as follows\cite{kastor}
\begin{equation}\label{1.12}
\begin{split}
\delta M&=T\delta S+V\delta P+\Phi\delta Q+\Omega\delta J\\
&=\frac{\kappa}{2\pi}\delta S+V \delta P+\Phi\delta Q+\Omega\delta J\\
&=\frac{\kappa}{8\pi G}\delta A-\frac{V}{8\pi G}\delta \Lambda+\Phi\delta Q+\Omega\delta J
\end{split}
\end{equation} 
where we have used eq.(s) (\ref{1.8},~\ref{1.10},~\ref{1.11}) to obtain the final expression in eq.(\ref{1.12}).

\noindent Holographic interpretation of the above first law of thermodynamics is shown to have some issues \cite{kostar2,Dolan1,Kastor,Zhang,Zhang2,Dolan22,McCarthy,JohnMarSve}. The pressure-volume term (corresponding to the variation of the cosmological constant) in the first law of thermodynamics of the bulk is shown to have two terms in the first law of thermodynamics at the boundary CFT. These are the central charge of the boundary CFT (and its corresponding conjugate variable)\cite{kostar2}, and the thermodynamic pressure (and its conjugate variable, volume) of the boundary CFT (because change in the AdS radius will correspond to change in the boundary radius\cite{karch}).  The way to deal with this problem is to invoke the form of the central charge from the $AdS/CFT$ dictionary. In this way we can avoid the ambiguity by varying both the Newton's gravitational constant and the $AdS$ radius( or varying the cosmological constant). The $AdS/CFT$ dictionary relates the central charge $C$ to the $AdS$ radius $l$ as \cite{karch}
\begin{equation}\label{1.13}
C=k\frac{l^{D-2}}{16\pi G}
\end{equation}
where  the $k$ factor depends on the details of the system at the boundary. From the above expression of $C$,  we see that in order to keep C fixed, we need to vary $G$ as well since $l$ is varying\cite{mann1}. In case of a Born-Infeld $AdS$ black hole, there is an additional parameter which is the Born-Infeld parameter $b$. If we look at the setup carefully, we shall find that all the parameters which are being varied are dimensionful. The Born-Infeld parameter also has a dimension. Hence, we treat it as a thermodynamic variable. We consider the black hole mass $M$ to be a function of area ($A$), charge ($Q$), angular momentum ($J$), Newtonian gravitational constant ($G$), cosmological constant ($\Lambda$), and the Born-Infeld parameter ($b$). Hence, we write the black hole mass as 
\begin{equation}\label{1.14}
M\equiv M(A,Q,J,G,\Lambda ,b)~.
\end{equation}
The variation in the mass can be written as
\begin{equation}\label{1.15}
\delta M=\frac{\partial M}{\partial A}\delta A+\frac{\partial M}{\partial Q}\delta Q+\frac{\partial M}{\partial J}\delta J+\frac{\partial M}{\partial G}\delta G+\frac{\partial M}{\partial\Lambda}\delta\Lambda+\frac{\partial M}{\partial b}\delta b.
\end{equation}
We shall define $\frac{\partial M}{\partial b}=\mathcal{B}$ and $G\frac{\partial M}{\partial G}=-\xi$. Note that the conjugate variable $\mathcal{B}$ to the Born-Infeld parameter $b$ has been termed as \textit{Born-Infeld vacuum polarization} in \cite{gunasekaran}. From eq.(\ref{1.12}), we find that the conjugate variables of $A,~\Lambda,~Q$ and $J$ are $\frac{\kappa}{8\pi G}$, $-\frac{V}{8\pi G}$, $\Phi$, and $\Omega$.
With these definitions, we can recast eq.(\ref{1.15}) as
\begin{equation}\label{1.16}
\delta M=\frac{\kappa}{8\pi G}\delta A+\Phi\delta Q+\Omega\delta J-\xi\frac{\delta G}{G}-\frac{V}{8\pi G}\delta\Lambda+\mathcal{B}\delta b~.
\end{equation}
Our main aim now is to compute the coefficient $\xi$ of $\delta G$ in the above expression. For that we now make use of a modified mass term as suggested in \cite{mann1}

\begin{equation}\label{1.17}
GM=\mathcal{M}(A,\sqrt{G}Q,\Lambda,G J, b)~.
\end{equation}
Taking differential of the both sides of the above relation, we get
\begin{equation}\label{1.18}
\begin{split}
\delta(GM)&=\frac{\partial \mathcal{M}}{\partial A}\delta A+\frac{\partial\mathcal{M}}{\partial(\sqrt{G}Q)}\delta (\sqrt{G}Q)+\frac{\partial\mathcal{M}}{\partial\Lambda}\delta\Lambda\\
&+\frac{\partial\mathcal{M}}{\partial(GJ)}\delta(GJ)+\frac{\delta\mathcal{M}}{\delta b}\delta b\\
\implies G\delta M&=-M\delta G+\frac{\partial \mathcal{M}}{\partial A}\delta A+\sqrt{G}\frac{\partial\mathcal{M}}{\partial (\sqrt{G}Q)}\delta Q\\
&+\frac{Q}{2\sqrt{G}}\frac{\partial\mathcal{M}}{\partial (\sqrt{G}Q)}\delta G+J\frac{\partial\mathcal{M}}{\partial (GJ)}\delta G\\
&+G\frac{\partial\mathcal{M}}{\partial (GJ)}\delta J+\frac{\partial\mathcal{M}}{\partial\Lambda}\delta\Lambda+\frac{\partial\mathcal{M}}{\partial b}\delta b\\
\implies \delta M&=\frac{1}{G}\frac{\partial \mathcal{M}}{\partial A}\delta A+\frac{1}{\sqrt{G}}\frac{\partial \mathcal{M}}{\partial (\sqrt{G}Q)}\delta Q+\frac{\partial\mathcal{M}}{\partial (GJ)}\delta J\\
&+\frac{1}{G}\left(-M+\frac{Q}{2\sqrt{G}}\frac{\partial\mathcal{M}}{\partial (\sqrt{G}Q)}+J\frac{\partial \mathcal{M}}{\partial(GJ)}\right)\delta G\\
&+\frac{\partial \mathcal{M}}{\partial \Lambda}\delta \Lambda+\frac{\partial\mathcal{M}}{\partial b}\delta b ~.
\end{split} 
\end{equation}
Now comparing eq.(\ref{1.18}) with eq.(\ref{1.16}), we obtain the following results 
\begin{equation}\label{1.19}
\begin{split}
\frac{\partial\mathcal{M}}{\partial A}&=\frac{\kappa}{8\pi}~, ~\frac{1}{\sqrt{G}}\frac{\partial \mathcal{M}}{\partial (\sqrt{G}Q)}=\Phi~,~\frac{\partial \mathcal{M}}{\partial (GJ)}=\Omega~,\\
\frac{\partial \mathcal{M}}{\partial \Lambda}&=-\frac{V}{8\pi G}~,~\frac{\partial\mathcal{M}}{\partial b}=\mathcal{B}~
\end{split}
\end{equation}
and the conjugate variable to $G$ is given as follows
\begin{equation}\label{1.20}
\xi=M-\frac{Q}{2\sqrt{G}}\frac{\partial\mathcal{M}}{\partial (\sqrt{G}Q)}-J\frac{\partial \mathcal{M}}{\partial(GJ)}~.
\end{equation}
Using the relations from eq.(\ref{1.19}) in eq.(\ref{1.20}), we obtain the following form of $\xi$
\begin{equation}\label{1.21}
\xi=M-\frac{Q\Phi}{2}-\Omega J~.
\end{equation}
This is the form of the conjugate variable to $G$ in terms of the other black hole parameters.
  
\noindent Now taking the differential of eq.(\ref{1.13}) and dividing by $C$, we get
\begin{equation}\label{1.22}
\frac{\delta C}{C}=-\frac{\delta G}{G}+(D-2)\frac{\delta l}{l}~.
\end{equation} 
In eq.(\ref{1.22}), we replace $\frac{\delta l}{l}$ by using the forms of $P$ and $\Lambda$ in eq.(s)(\ref{1.8},~\ref{1.9}) as follows
\begin{equation}\label{1.22a}
\frac{\delta l}{l}=-\frac{\delta G}{2G}-\frac{\delta P}{2P}.
\end{equation}
Eq.(\ref{1.22a}) enables us to recast eq.(\ref{1.22}) equation in the form
\begin{equation}\label{1.23}
\frac{\delta G}{G}=-\frac{2}{D}\frac{\delta C}{C}-\frac{(D-2)}{D}\frac{\delta P}{P}~.
\end{equation}
Using eq.(s)(\ref{1.19},~\ref{1.20},~\ref{1.21},~\ref{1.23}) in eq.(\ref{1.18}), we get the form of $\delta M$ to be
\begin{equation}\label{1.24}
\begin{split}
\delta M&=\frac{\kappa}{8\pi G}\delta A+\Phi\delta Q+\Omega\delta J-\frac{V}{8\pi G}\delta\Lambda+\mathcal{B}\delta b\\
&+\frac{2\xi}{DC}\delta C+\frac{(D-2)}{D}\xi\frac{\delta P}{P}~.
\end{split}
\end{equation}
Using eq.(s)(\ref{1.8},~\ref{1.10},~\ref{1.11},~\ref{1.23}), we can now rewrite eq.(\ref{1.24}) as follows
\begin{equation}\label{1.25}
\begin{split}
\delta M&=T\delta S+\phi\delta Q+\Omega\delta J+\left[\frac{2\xi}{DC}-\frac{2(TS+PV)}{DC}\right]\delta C\\
&+\left[V+\frac{D-2}{DP}\xi-\frac{D-2}{DP}(TS+PV)\right]\delta P+\mathcal{B}\delta b\\
&=T\delta S+\phi\delta Q+\Omega\delta J+\mathcal{B}\delta b+V_\mathcal{C}\delta P+\mu_\mathcal{C}\delta C
\end{split}
\end{equation}
where
\begin{align}
V_\mathcal{C}&=V+\frac{D-2}{DP}\xi-\frac{D-2}{DP}(TS+PV)\label{1.26}~,\\
\mu_\mathcal{C}&=\frac{2\xi}{DC}-\frac{2(TS+PV)}{DC}\label{1.27}~.
\end{align}
Here $V_\mathcal{C}$ and $\mu_\mathcal{C}$ are the new effective  thermodynamic volume and chemical potential. Eq.(\ref{1.25}) is the desired mixed form of the first law of thermodynamics for a Born-Infeld $AdS$ black hole. This is one of the main findings in our paper. The important point to note in eq.(\ref{1.25}) is that $\delta C$ can be set to zero since both $l$ and $G$ are varying. This means keeping the boundary CFT intact. This allows us to study the thermodynamics of the bulk with a fixed central charge in the boundary theory. The reason so as to why this is called a mixed form of the first law is that it contains both the bulk as well as boundary variables.

\subsection{Extended black hole thermodynamics and the Smarr relation}
In this subsection, we want to calculate V in terms of the mass M, charge Q and Born-Infeld parameter b of the black hole. For that, we write down the first law for Born-Infeld $AdS$ black holes as \cite{gunasekaran}
\begin{equation}\label{1.28}
\delta M=T\delta S+V\delta P+\Phi \delta Q+\Omega\delta J+\mathcal{B}\delta b~.
\end{equation}
Here, $\Phi$, $\Omega$ and $\mathcal{B}$ are conjugate to $Q$, $J$ and $b$ respectively, where 
\begin{align}\label{1.29}
\Phi=\frac{\delta M}{\delta Q}~,~~\Omega=\frac{\delta M}{\delta J}~,~~\mathcal{B}=\frac{\delta M}{\delta b}~.
\end{align}
$\mathcal{B}$ has been called the Born-Infeld vacuum polarization and its inclusion in first law as well as in Smarr relation has been thoroughly discussed in \cite{Rash,Breton,Huan,ext2}. Here, in our consideration $M=M(S,P,b,Q,J)$. 
In generalized $D$-dimensions, $M,S,P,b,Q,J$ have the following dimensions (in terms of $L$)
\begin{equation}\label{1.30}
\begin{split}
[M]&=L^{D-3},~[S]=L^{D-2},~[P]=L^{-2}~,\\
[b]&=L^{-1},~[Q]=L^{D-3},~[J]=L^{D-2}~.
\end{split}
\end{equation}
\noindent Using Euler's theorem of quasi-homogeneous functions, we get
\begin{equation}\label{1.31}
\begin{split}
(D-3)M&=(D-2)S\frac{\delta M}{\delta S}-b\frac{\delta M}{\delta b}-2P\frac{\delta M}{\delta P}\\&+(D-3)Q\frac{\delta M}{\delta Q}+(D-2)J\frac{\delta M}{\delta J}~.
\end{split}
\end{equation}
\noindent Using eq.(\ref{1.28}), we can write $\frac{\partial M}{\partial S}=T$, $\frac{\partial M}{\partial Q}=\Phi$, $\frac{\partial M}{\partial P}=V$, $\frac{\delta M}{\delta J}=\Omega$ and $\frac{\partial M}{\partial b}=\mathcal{B}$. Substituting them back in eq.(\ref{1.31}), we get the Smarr formula as follows 
\begin{equation}\label{1.32}
(D-3)M=(D-2)TS-\mathcal{B} b-2PV+(D-3)\Phi Q+(D-2)\Omega J~.
\end{equation}
\noindent From eq.(\ref{1.32}), we can obtain the form for the black hole volume V to be
\begin{equation}\label{1.33}
\begin{split}
V&=\frac{D-2}{2P}TS-\frac{\mathcal{B}b}{2P}+\frac{D-3}{2P}\Phi Q-\frac{D-3}{2P}M+\frac{D-2}{2P}\Omega J\\
&=\frac{D-3}{2P}\left(\frac{D-2}{D-3}(TS+\Omega J)+\Phi Q-M-\frac{1}{D-3}\mathcal{B}b\right)~.
\end{split}
\end{equation}
\subsection{Forms of the modified thermodynamic variables}
Replacing the form of $V$ from eq.(\ref{1.33}) in eq.(\ref{1.26}), we get
\begin{equation}\label{1.34}
V_\mathcal{C}=\frac{2M-2\mathcal{B}b+(D-4)Q\Phi}{2DP}~.
\end{equation}  
Similarly, we can write the chemical potential in terms of $V$ and $V_\mathcal{C}$ as
\begin{equation}\label{1.35}
\mu_\mathcal{C}=\frac{2P}{C(D-2)}(V_\mathcal{C}-V)~.
\end{equation}
In $(3+1)$-dimensions, that is, $D=4$, eq.(s)(\ref{1.34},\ref{1.35}) reduces to the following forms
\begin{equation}\label{1.36}
\begin{split}
V_\mathcal{C}&=\frac{M-\mathcal{B}b}{DP}
\end{split}
\end{equation}
\begin{equation*}
\begin{split}
V_\mathcal{C}&=\frac{4\pi r_+^3}{3}+\frac{4\pi l^2 r_+}{3}-\frac{8\pi b^2 l^2r_+^3}{9}+\frac{8\pi b^2 l^2r_+^3}{9}\sqrt{1+\frac{GQ^2}{b^2r_+^4}}\\
&+\frac{8\pi GQ^2l^2}{9r_+}\,_2F_1\left[\frac{1}{4},\frac{1}{2},\frac{5}{4},-\frac{GQ^2}{b^2r_+^4}\right]~,
\end{split}
\end{equation*}
\begin{equation}\label{1.37}
\begin{split}
\mu_\mathcal{C}&=\frac{P(V_\mathcal{C}-V)}{C}\\
&=\frac{4\pi r_+}{kl^2}-\frac{8b^2\pi r_+^3}{3kl^2}+\frac{8b^2\pi r_+^3}{3kl^2}\sqrt{1+\frac{GQ^2}{b^2r_+^4}}\\
&+\frac{8\pi GQ^2}{3kl^2r_+}\,_2F_1\left[\frac{1}{4},\frac{1}{2},\frac{5}{4},-\frac{GQ^2}{b^2r_+^4}\right]
\end{split}
\end{equation}
where we have used eq.(\ref{1.6}), 
\begin{equation}\label{1.37a}
P=\frac{3}{8\pi l^2G},~C=\frac{kl^2}{16\pi G},~\Phi=\frac{Q}{r_+}\,_2F_1\left[\frac{1}{4},\frac{1}{2},\frac{5}{4},-\frac{GQ^2}{b^2r_+^4}\right]~,
\end{equation}
and
\begin{equation}\label{1.37b}
\begin{split}
 \mathcal{B}&=\frac{\partial M}{\partial b}\\&=\frac{2br_+^3}{3G}\left[1-\sqrt{1+\frac{GQ^2}{b^2r_+^4}}\right]+\frac{Q^2}{3br_+}\,_2F_1\left[\frac{1}{4},\frac{1}{2},\frac{5}{4},-\frac{GQ^2}{b^2r_+^4}\right]
\end{split}
\end{equation} 
 to obtain the final forms of $V_\mathcal{C}$ and $\mu_\mathcal{C}$. $V_\mathcal{C}$ and $\mu_\mathcal{C}$ are the new thermodynamic variables for the Born-Infeld $AdS$ black hole. Interestingly, we find that the new volume $V_\mathcal{C}$ depends on the Born-Infeld parameter which enters through the expression for $V$ obtained from the Smarr formula. Our next goal is to understand the effects of the Born-Infeld term on the critical behaviour of the black hole and the possible phase transition structure.
\section{Breaking of universal nature of the central charge}
\noindent Now we shall calculate the critical  value of the central charge. In order to find this value, we will use the following two equations\cite{mann1}
\begin{align}
\dfrac{\partial T}{\partial r_+}&=0~,\label{1.39}\\
\dfrac{\partial^2T}{\partial r_+^2}&=0\label{1.40}~.
\end{align}
Using eq.(\ref{1.7}), we get the following equation from eq.(\ref{1.40})
\begin{equation}\label{1.41}
1-\frac{6GQ^2}{r_{+(c)}^2\left(1+\frac{GQ^2}{b^2r_{+(c)}^4}\right)^{\frac{1}{2}}}+\frac{4G^2Q^4}{b^2r_{+(c)}^4\left(1+\frac{GQ^2}{b^2r_{+(c)}^4}\right)^{\frac{3}{2}}}=0~.
\end{equation}
Here, $c$ in the subscript of $r_{+(c)}$ is denoting the critical value. 
Eq.(\ref{1.41}) upto $\mathcal{O}(1/b^2)$ takes the form as

\begin{equation}\label{1.41a}
r_{+(c)}^6-6GQ^2r_{+(c)}^4+\frac{7G^2Q^4}{b^2}=0~.
\end{equation}
Now we proceed to obtain the solution of eq.(\ref{1.41a}) by approximating it upto $\mathcal{O}(1/b^2)$. We take a perturbative approach to obtain the critical value of $r_{+(c)}$. We propose a solution of the form for $r_{+(c)}$ (the critical value of $r_+$) upto $\mathcal{O}(1/b^2)$ as follows
\begin{equation}\label{1.41b}
r_{+(c)}\cong r_+^{(0)}+\frac{r_+^{(1)}}{b^2}~.
\end{equation}
Using eq.(\ref{1.41b}) in eq.(\ref{1.41a}), we obtain the forms of $r_+^{(0)}$ and $r_+^{(1)}$ as follows
\begin{equation}\label{1.41c}
r_+^{(0)}=\sqrt{6G}Q~,~~r_+^{(1)}=-\frac{7}{72\sqrt{6G}Q}~.
\end{equation} 
Eq.(\ref{1.41c}) gives the form of the critical value of $r_{+(c)}$ upto $\mathcal{O}(1/b^2)$ as 
\begin{equation}\label{1.42}
r_{+(c)}\cong \sqrt{6G}Q-\dfrac{7}{72\sqrt{6G}Q b^2}~.
\end{equation}
Now, eq.(\ref{1.39}) can be recast in the following form
\begin{equation}\label{1.42a}
\begin{split}
&-\frac{1}{r_{+(c)}^2}+\frac{3}{l^2}+2b^2-2b^2\sqrt{1+\frac{GQ^2}{b^2r_{+(c)}^4}}\\&+\frac{4GQ^2}{r_{+(c)}^4\sqrt{1+\frac{GQ^2}{b^2r_{+(c)}^4}}}=0~.
\end{split}
\end{equation}
Eq.(\ref{1.42a}) upto $\mathcal{O}(1/b^2)$ can be recast in the following form
\begin{equation}\label{1.42b}
-\frac{1}{r_{+(c)}^2}+\frac{3}{l^2}-\frac{7G^2Q^4}{4b^2r_{+(c)}^8}+\frac{3GQ^2}{r_{+(c)}^4}=0~.
\end{equation}
Substituting eq.(\ref{1.42}) in eq.(\ref{1.42b}), we obtain the critical value of the $AdS$ radius upto $\mathcal{O}(1/b^2)$ as
\begin{equation}\label{1.43}
l_c\cong 6\sqrt{G}Q\left(1-\dfrac{7}{864GQ^2b^2}\right)~.
\end{equation}
Putting the values of $r_{+(c)}$ and $l_c$ in eq.(\ref{1.7}), we obtain the critical value of the temperature upto $\mathcal{O}(1/b^2)$ as
\begin{equation}\label{1.44}
T_c=\dfrac{1}{3\sqrt{6G}\pi Q }+\dfrac{1}{432\pi \sqrt{6G}GQ^3b^2}~.
\end{equation}
All the results reduce to the Reissner-Nordstr$\ddot{\text{o}}$m case in the limit ($b\rightarrow\infty$)\cite{mann1}. We can obtain the value of the critical charge by using the form of $l_\mathcal{C}$ from eq.(\ref{1.43}) in eq.(\ref{1.13}) upto $\mathcal{O}(1/b^2)$ as follows
\begin{equation}\label{1.45}
\begin{split}
C_c&=k\dfrac{l_c^2}{16\pi G}\\
\implies C_c&\cong\dfrac{9kQ^2}{4\pi}-\dfrac{7k}{16\pi G b^2}~.
\end{split}
\end{equation}
Eq.(\ref{1.45}) gives the form of the critical charge in terms of $Q,~G$ and $b$. We will now proceed to calculate the form of the critical charge in terms of $Q,~b$ and the critical pressure $P_c$. This  form is needed to avoid using values of $G$ while calculating the critical values of the central charge $C_c$ for different values of the non-linear parameter $b$ (Table \ref{tab1}).

\noindent In four spacetime dimensions and for $l=l_c$ we can recast eq.(s)(\ref{1.8},\ref{1.9}) as
\begin{equation}\label{1.45A}
P_c=\frac{3}{8\pi G l_c^2}~.
\end{equation}
Again, replacing $l_c^2$ by $\frac{16\pi G C_c}{k}$ from the first line of eq.(\ref{1.45}) in the above equation, we obtain (after taking a square root)
\begin{equation}
G=\sqrt{\frac{3k}{128\pi^2C_c P_c}}
\end{equation}
where $P_c$ is the pressure corresponding to the critical value of the central charge. Using the form of the gravitational constant in eq.(\ref{1.45}), we can recast eq.(\ref{1.45}) in the following form
\begin{equation}\label{1.46}
C_c=\frac{9kQ^2}{4\pi}-\frac{7\sqrt{kP_c}}{\sqrt{6}b^2}C_c^{\frac{1}{2}}~.
\end{equation}
Solving eq.(\ref{1.46}) upto $\mathcal{O}(1/b^2)$, the form of the critical charge in terms of $P_c$, the Born-Infeld parameter $b$, black hole charge $Q$ and $k$ is given by
\begin{equation}\label{1.46c}
C_c\cong\frac{9kQ^2}{4\pi}-\frac{7kQ}{4b^2}\sqrt{\frac{6P_c}{\pi}}~.
\end{equation}
In case of Reissner-Nordstr$\ddot{\text{o}}$m black hole, the critical value of the central charge was obtained as\cite{mann1}
\begin{equation}\label{1.45a}
C_c=\dfrac{9kQ^2}{4\pi}~.
\end{equation}
In the above equation, we see that the central charge $C_c$ is universal in nature. In case of a Born-Infeld AdS black hole, we observe from eq.(\ref{1.46c}) that the universal nature of the central charge is now broken due to the presence of the Born-Infeld parameter and the critical-pressure term $P_c$ in it. Dependence of the critical value of the central charge on the Born-Infeld parameter is given in Table(\ref{tab1}). To observe this dependence, we used $Q=1$,~$k=16\pi$, and $P_c=15$. 
\begin{table}
\centering
\caption{Dependence of $C_c$ on $b$}
\medskip
\begin{tabular}{|c||c|c|c|c|c|c|}
\hline
$b$ &$10$&$15$&$20$&$30$&$100$&$\infty$\\
\hline
$C_c$&$31.29$&$33.91$&$34.82$&$35.48$&$35.95$&$36$
\\\hline
\end{tabular}
\label{tab1}
\end{table} 
For the above values of the parameters, $C_c$ for a Reissner-Nordstr$\ddot{\text{o}}$m $AdS$ black hole has the value $36$. From Table(\ref{tab1}), we observe that $C_c$ value for the Born-Infeld case approaches $36$ as we increase the parameter value $b$ without changing the pressure term ($P$). In the limit $b\rightarrow\infty$, the critical value of the central charge approaches the Reissner-Nordstr$\ddot{\text{o}}$m black hole result.
\section{Phase transition structure}
\begin{figure}[ht!]
\centering
\includegraphics[scale=0.28]{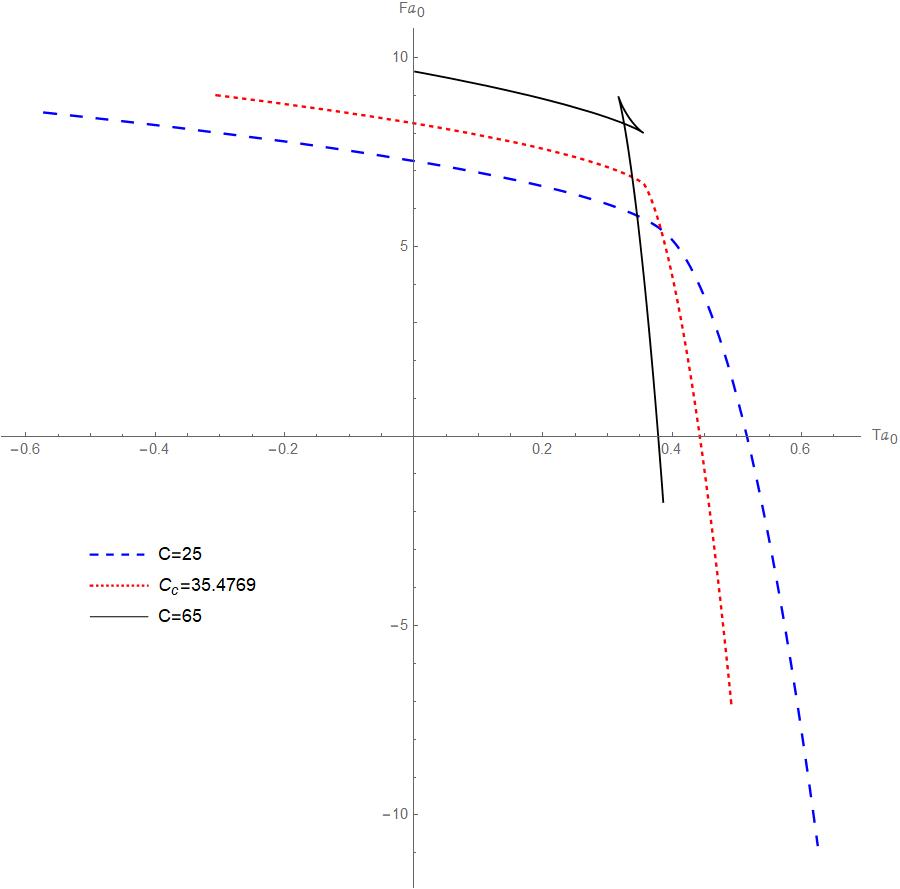}
\caption{Free energy vs Temperature: \
\textit{Parameters} Q=1; k=16$\pi$; P=15; b=30 \textit{($a_0$ is a scaling parameter)}: a)  $C=25$ (\textit{Dashed line}), b) $C=C_c=35.4769$ (\textit{Dotted line}) , c) $C=60$ (\textit{Solid line}) } 
\label{f11}
\end{figure}

 In this section, we will try to investigate the phase transition structure for a Born-Infeld $AdS$ black hole. The free energy of the black hole can be computed using eq.(s)(\ref{1.6},\ref{1.7}) along with the use of the Bekenstein-Hawking entropy as follows
\begin{equation}
\begin{split}\label{1.38}
F&=M-TS\\
&=\dfrac{r_+}{4G}-\dfrac{r_+^3}{4G~l^2}\dfrac{b^2r_+^3}{6G}\left(1-\sqrt{1+\dfrac{GQ^2}{b^2r_+^4}}\right)\\&+\dfrac{2Q^2}{3r_+}\,_2F_1\left[\frac{1}{4},\frac{1}{2},\frac{5}{4},-\dfrac{GQ^2}{b^2r_+^4}\right]~.
\end{split}
\end{equation}
Here, in principle, we can write $F\equiv F(T,P,Q,C,b)$. The behaviour of the free energy of the black hole with its temperature is plotted in Figure(\ref{f11}) for different values of the central charge. We observe that above a critical value of the central charge there is a swallowtail behaviour of the free energy of the black hole with respect to the change in the black hole temperature for fixed value of the pressure. Below this critical value of the central charge ($C_\mathcal{C}=35.4769$ in Figure(\ref{f11})), we observe a smooth curve for the free energy indicating no swallowtail behaviour. Critical analysis of Born-Infeld $AdS$ black holes in extended thermodynamics have been discussed in \cite{gunasekaran}. The importance of our work lies in the fact that it deals with the case where the contribution of the central charge from the CFT side  plays an important role which is why it is called a mixed thermodynamic behaviour.
\begin{figure}[ht!]
\centering
\includegraphics[scale=0.40]{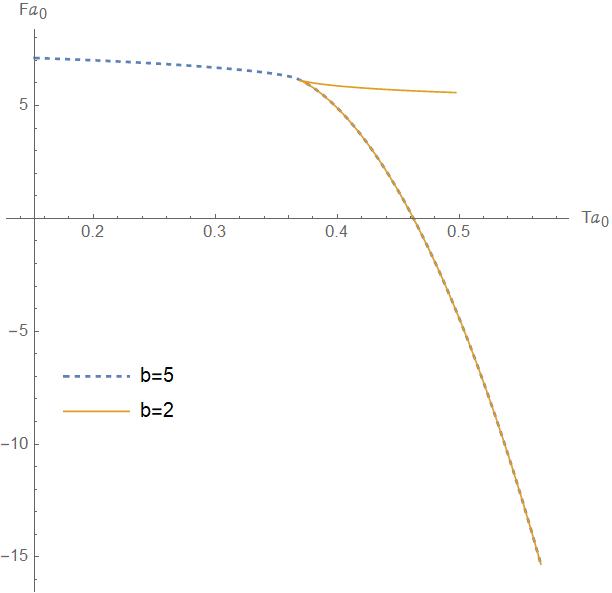}
\caption{Free energy vs Temperature: \
\textit{Parameters} Q=1; k=16$\pi$; P=15; C=30 \textit{($a_0$ is a scaling parameter)}: a)  $b=5$ (\textit{Dashed line}), b) $b=2$ (\textit{Solid line})} 
\label{f22}
\end{figure}
\begin{figure}[ht!]
\centering
\includegraphics[scale=0.40]{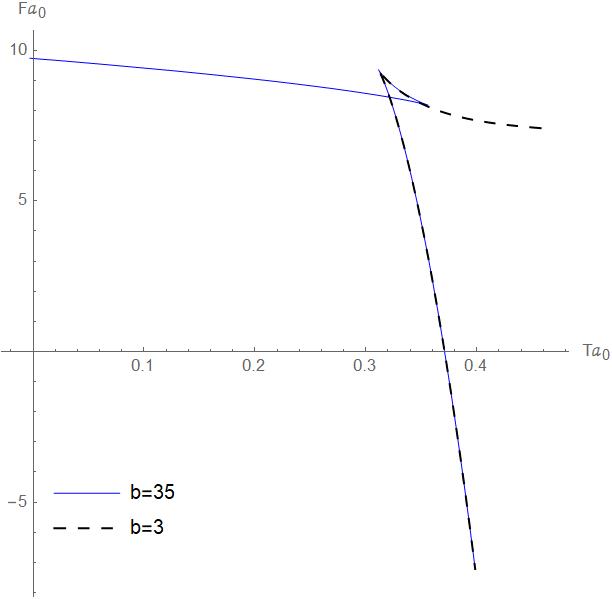}
\caption{Free energy vs Temperature: \
\textit{Parameters} Q=1; k=16$\pi$; P=15; C=70 \textit{($a_0$ is a scaling parameter)}: a)  $b=30$ (\textit{Dashed line}), b) $b=3$ (\textit{Solid line})} 
\label{f23}
\end{figure}


\noindent In Figure(\ref{f11}), we will investigate the behaviour of the free energy with respect to change in the temperature for various central charge values. As opposed to \cite{mann1}, the critical value of the central charge has a direct dependence on the pressure term as can be seen from eq.(\ref{1.46c}). Now from Figure (\ref{f11}), we observe that for the value of the central charge, $C=25$ (which is less than the critical value,  $C_c=35.4769$), the free energy  curve is smooth with respect to change in the temperature. For curves with the value of the central charge, $C\geq C_c$ (for fixed $P$ and $b$ values), the free energy curve no longer remains smooth. Instead we observe a swallowtail behaviour indicating first order small to large black hole phase transition. With increasing temperature, we now observe a drop in the free energy below the $F=0$ line. We also observe that for the value of the central charge below and above the critical value, the curves meet the $F=0$ line at respectively higher and lower values of the temperature. Hence, we observe for a Born-Infeld $AdS$ black hole, there exists a critical value of the central charge for fixed values of the pressure term and Born-Infeld parameter above which there is a small to large black hole phase transition.

\noindent For the next part of our analysis, we concentrate on the dependence of the critical value of the central charge on the Born-Infeld parameter $b$. We have plotted free energy with respect to the temperature for different values of the Born-Infeld parameter $b$ with fixed values of the central charge. In Figures(\ref{f22},\ref{f23}), we observe that for a substantially low value of the Born-Infeld parameter ($b=2$ for Figure(\ref{f22}) and $b=3$ for Figure(\ref{f23})), the free energy curve does not exist below a certain temperature, indicating the non-existence of any black hole in that region. Above this temperature value ($Ta_0\sim 0.36$ $(b=2)$ in Figure(\ref{f22}) and $Ta_0\sim 0.32$ $(b=3)$ in Figure(\ref{f23})), we observe that there are two branches of black hole solutions. However, the upper one is unstable (since the free energy is higher in this branch) and the lower one describes large black hole solutions which are physical.  This indicates  that for sufficiently small $b$, there exists a no black hole region and a branch of large black hole solutions. These features are qualitatively same with the Reissner-Nordstr$\ddot{\text{o}}$m $AdS$ black hole \cite{mann1} but Born-Infeld parameter plays a crucial role in the phase transition structure through its dependence on the critical value of the central charge.
\section{Conclusion}
In this work, we have studied the thermodynamics of a Born-Infeld AdS black hole in ($3+1$)-dimensions with variable $AdS$ radius, Newton's gravitational constant and the Born-Infeld parameter. Starting with the derivation of thermodynamic quantities of the black hole, we expressed the first law of thermodynamics in terms of the usual thermodynamical variables along with the central charge and the Born-Infeld parameter. The mixed from of the first law is derived in general $D$-spacetime dimensions with all thermodynamic variables. This modified first law for the Born-Infeld $AdS$ black hole system is called mixed because it has contributions from both the bulk as well as the boundary field theory variables. However, later we have reduced the discussion to four spacetime dimensions. Because of the existence of the Born-Infeld parameter in the Smarr relation, the thermodynamic variables $V_\mathcal{C}$ and $\mu_\mathcal{C}$ get modified by the same. Also, this modification leads to a change in the critical value of the central charge which again depends on the Born-Infeld parameter including a pressure dependence as well. This breaks the universal behaviour of the central charge observed in an earlier study\cite{mann1}. This is one of the most important results in our paper. From the free energy versus temperature plots, we observe that the phase structure undergoes crucial change due to the inclusion of the Born-Infeld parameter as a thermodynamical variable. We also observe that if the Born-Infeld parameter is decreased enough, there will be a certain critical temperature below which black holes do not exist. It implies that the inclusion of the electromagnetic non-linearity results in some significant change in the overall free energy behaviour of the black hole with respect to the change in the Hawking temperature.

\section*{Acknowledgement}
We thank the anonymous referee for his constructive comments. 

\end{document}